\documentclass[%
 reprint,
superscriptaddress,
 amsmath,amssymb,
 aps,
 pre,
]{revtex4-2}
\usepackage{balance}
\usepackage{graphicx}
\usepackage{dcolumn}
\usepackage{bm}
\usepackage{xcolor}
\usepackage[utf8]{inputenc}

\begin{document}

\preprint{APS/123-QED}

\title{High-order semi-Lagrangian kinetic scheme for compressible turbulence}

\author{Dominik Wilde}
\affiliation{
 Department of Mechanical Engineering, University of Siegen, Paul-Bonatz-Straße 9-11, 57076 Siegen-Weidenau, Germany}%
\affiliation{
Institute of Technology, Resource and Energy-efficient Engineering (TREE), Bonn-Rhein-Sieg University of Applied Sciences,
Grantham-Allee 20, 53757 Sankt Augustin, Germany
}

\author{Andreas Krämer}%
 \affiliation{
Department of Mathematics and Computer Science, Freie Universität Berlin, Arnimallee 6, 14195 Berlin, Germany
}%
 \author{Dirk Reith}
\affiliation{
Institute of Technology, Resource and Energy-efficient Engineering (TREE), Bonn-Rhein-Sieg University of Applied Sciences,
Grantham-Allee 20, 53757 Sankt Augustin, Germany
}
\affiliation{Fraunhofer Institute for Algorithms and Scientific Computing (SCAI), Schloss Birlinghoven, 53754 Sankt Augustin, Germany}

\author{Holger Foysi}

\affiliation{
 Department of Mechanical Engineering, University of Siegen, Paul-Bonatz-Straße 9-11, 57076 Siegen-Weidenau, Germany}%

\date{\today}

\begin{abstract}
Turbulent compressible flows are traditionally simulated using explicit time integrators applied to discretized versions of the Navier-Stokes equations. However, the associated Courant-Friedrichs-Lewy condition severely restricts the maximum time step size. Exploiting the Lagrangian nature of the Boltzmann equation's material derivative, we now introduce a feasible three-dimensional semi-Lagrangian lattice Boltzmann method (SLLBM), which circumvents this restriction. While many lattice Boltzmann methods for compressible flows were restricted to two dimensions due to the enormous number of discrete velocities in three dimensions, the SLLBM uses only 45 discrete velocities. Based on compressible Taylor-Green vortex simulations we show that the new method accurately captures shocks or shocklets as well as turbulence in 3D without utilizing additional filtering or stabilizing techniques, even when the time step sizes are up to two orders of magnitude larger compared to simulations in the literature. Our new method therefore enables researchers for the first time to study compressible turbulent flows by a fully explicit scheme, whose range of admissible time step sizes is only dictated by physics, while being decoupled from the spatial discretization. 
\end{abstract}

\maketitle

\section{Introduction} 
One major challenge in fluid dynamics is the study of compressible turbulent flows, involving intrinsic as well as variable density compressibility effects \cite{Lele1994,Chassaing:2002,Veynante:2002,Smits2006,Garnier2009,Pirozzoli2011,GatskiBook}. Applications range from aviation \cite{Sharman2016} or astrophysics \cite{Low2004} to the investigation of canonical flows like boundary layers \cite{Fernholz:1976}, channel flow \cite{Huang:1995,Foysi:2004,Gosh:2010}, mixing layers \cite{Bradshaw:1977,Bogdanoff1983,Vreman:1996,Foysi:2010b,Matsuno2020}, jets and aeroacoustics \cite{Colonius:2004,Bodony:2006,Foysi:2010a} or shock-turbulence interaction \cite{Lele_2009}, to only mention a few considering the vast literature available. These flows feature both solenoidal and dilatational structures, which constantly interact and possibly cause shock waves \cite{Lee1991,Garnier2009}. 

Numerical simulations have become an indispensable tool to understand their physics, and many studies exploring compressible turbulent flows have been conducted using high-order compact finite difference, optimized dispersion-relation preserving schemes \cite{Lele1992,TamWebb,Adams1996,Ekaterinaris1999,Sengupta2003,Colonius:2004,Bogey2004,Wang2010} for the spatial derivatives, often combined with low-dispersion-dissipation Runge-Kutta schemes for time-integration \cite{Hu:1996,Colonius:2004,Berland:2006}. Although these methods provide accurate results, the time steps are generally small \cite{Kwatra2009}, because of the methods' Eulerian time derivatives, which describe how the variables of interest pass through fixed locations in the field. Thus, the admissible time step sizes are tightly linked to spatial resolution. This issue is for many discretizations linked to the Courant-Friedrichs-Lewy (CFL) condition,
\begin{equation} \label{eq:CFL}
    c{\delta_t}/{\delta_x} < \mathrm{CFL}_\mathrm{max},
\end{equation}

using linear stability theory, relating a characteristic velocity $c$  to the spatial and temporal discretization intervals $\delta_x$ and $\delta_t$, respectively (see \cite{Mueller:1990}, for example). Though implicit time integration schemes often provide larger stability domains, their application can be unfeasible for transient problems due to their computational cost.
Explicit time integration schemes with scheme-specific $\mathrm{CFL_{max}}$, by contrast, enforce small time steps $\delta_t$ for high flow velocities, typically occurring in many high-speed compressible flows. Another obvious way to circumvent the CFL condition in Eq. \eqref{eq:CFL} is to incorporate Lagrangian time derivatives, which track the motion of the variables of interest moving through the domain.\\ In practice, Semi-Lagrangian (SL) schemes are used instead, which provide a viable alternative to the discretization of Eulerian time derivatives.
SL schemes discretize the Lagrangian solution by  tracking the trajectories back in time. The prefix ``semi'' indicates that the trajectories' end points usually do not coincide with the simulation grid points, which requires application of an appropriate interpolation scheme. SL methods were successfully incorporated in algorithms solving the Navier-Stokes equations \cite{Xiu2001}, although tracking of the fluid trajectories was often found to be cumbersome, introducing additional errors \cite{Celledoni2016}. The major advantage when using SL schemes in kinetic theory is that the trajectories are \emph{linear}, resulting in cancellation of the tracking error. Consequently, SL schemes were both applied to the Vlasov equation \cite{Sonnendrucker1999, Crouseilles2010, Qiu2010} and to the Bhatnagar-Gross-Krook (BGK)-Boltzmann equation \cite{Russo2012,groppi2014high}. Recently, we introduced 
the semi-Lagrangian lattice Boltzmann method (SLLBM) \cite{Kramer2017,Kramer2020} for compressible flows \cite{Wilde2019}, which solves the lattice Boltzmann equation using a high-order SL streaming step.

In this article, we explore the capabilities of the SLLBM for three-dimensional compressible flows. Furthermore, we demonstrate that the SLLBM remains stable for time step sizes that exceed typical CFL constraints of Eulerian solvers by orders of magnitude. To yield a lean scheme, the SLLBM is combined with state-of-the-art cubature rules for the velocity discretization \cite{Stroud1973,Cools2003, Wilde2021}. This combination proves capable of modeling compressible turbulence with time steps that are at least one order of magnitude larger than in standard Eulerian methods and \emph{decouple} the spatial from the temporal discretization. 
\subsection{Background}
We start with a critical look at the more specialized lattice Boltzmann method (LBM) \cite{Kruger2016}. Despite the successes of the standard LBM in the computation of multiphase \cite{Chen2014}, particle-laden \cite{Wang2016}, thermal \cite{Peng2003}, or turbulent flows \cite{Geier2015, Dorschner2016}, compressible LBM \cite{Alexander1992} were overlooked for a relatively long time, but regained attraction during the last decade \cite{Frapolli2015, Frapolli2016,Frapolli2016a,Feng2016,Coreixas2017,Atif2018,Dorschner2018,Feng2019,Hosseini2019,Saadat2019,Saadat2020,Latt2020,Farag2020,Coreixas2020a}.
Let us recall the force-free BGK-Boltzmann equation 
\begin{equation}\label{eq:Boltzmann}
    \frac{\partial f }{\partial t} + \boldsymbol \xi \cdot \nabla f= -\frac{1}{\lambda}\left(f -f^\mathrm{eq}\right),
\end{equation} with the continuous distribution function $f$, the equilibrium distribution function $f^\mathrm{eq}$, the particle velocity $\boldsymbol{\xi}$, and the relaxation time $\lambda$. To discretize Eq. \eqref{eq:Boltzmann}, the original LBM is based on
three key principles. First, the equilibrium distribution function $f^\mathrm{eq}$ is polynomially expanded into a series of Hermite polynomials $\mathcal{H}^{(n)}$, with expansion coefficients being the equilibrium moments $\mathbf{a}^{(n)}_\mathrm{eq}$ \cite{Shan1998},\vspace{-.2cm}
\begin{equation}
\label{eq:equilibrium} f^{\mathrm{eq}}\approx  \omega(\boldsymbol{\xi}) \sum_{n=0}^N \frac{1}{n!}{\boldsymbol{ a}}^{(n)}_\mathrm{eq} : {\boldsymbol{\mathcal{H}}} ^{(n)},
\end{equation}
where $N$ is the expansion order and $\omega(\boldsymbol{\xi})$ the weight function. Since $\boldsymbol{a}^{(n)}_\mathrm{eq}$ and $\boldsymbol{\mathcal{H}}^{(n)}$ are symmetric tensors of rank $n$, the product involves contraction to all $D^n$ scalar components, depending on dimension $D$. Second, a Gauß-Hermite quadrature is applied to the unbounded velocity space of the Boltzmann equation, leading to discrete particle velocity sets \cite{Shan1998}. The moments are then found by the quadrature 
\begin{equation}\label{eq:quadrature}
    \boldsymbol{a}^{(n)} = \int_{\mathbb{R}^D} \omega(\boldsymbol{\xi}) \frac{f}{\omega(\boldsymbol{\xi})} \boldsymbol{\mathcal{H}}^{(n)}(\boldsymbol{\xi})d\boldsymbol{\xi} = \sum_{i=0}^{Q-1}
    f_i \boldsymbol{\mathcal{H}}^{(n)}(\boldsymbol{\xi}_i).
\end{equation}
with the weighted discrete distribution functions $f_i = w_i f(\boldsymbol{\xi}_i) / \omega(\boldsymbol{\xi}_i).$
The combination of $Q$ discrete particle velocities $\boldsymbol{\xi}_i$ and  weights $w_i$, the velocity set, is usually derived by the Gauß-product rule applied to a one-dimensional Gauß-Hermite quadrature.
Third, the discrete Boltzmann equation is integrated along characteristics with an inherent Lagrangian discretization of the Boltzmann equation's material derivative to obtain a stable numerical scheme and second-order temporal convergence \cite{He1998}.

Unfortunately, the LBM in its original formulation is mainly restricted to Cartesian grids and velocity discretizations that match the regular lattices. The customary ``D2Q9'' based on second-order expansion in Eq. \eqref{eq:equilibrium} is plagued by a cubic error being proportional to the Mach number. Consequently, compressible simulations either demand correction terms that annihilate the errors and restore Galilean invariance \cite{Prasianakis2007, Saadat2019} or higher-order discretizations of Eqs. \eqref{eq:equilibrium} and \eqref{eq:quadrature} which render abscissae that reside off-lattice.
Utilization of such velocity sets therefore requires an efficient off-lattice Boltzmann solver. Previous Eulerian off-lattice Boltzmann schemes \cite{Bardow2006,Lee2003,Min2011,Guo2013a, Guo2015}, like finite difference or finite volume LBM, would be suited  in principle. However, they sacrifice the Lagrangian time integration along characteristics. Moreover, their time step is severely restricted by a CFL condition, Eq. \eqref{eq:CFL}, with respect to the fast discrete particle velocities.

In contrast, the SLLBM preserves all of the aforementioned key principles of the LBM but it also decisively extends its capabilities. In previous works \cite{Kramer2017,Kramer2020,Wilde2019} we have shown that a high-order interpolation increases the spatial order of the method and nihilates mass losses. Also, we have demonstrated the unconditional stability of the advection step, when incorporating Gauß-Lobatto-Chebyshev nodes for the interpolation up to third order, and that the stability is practically not affected even with fourth order. The flexibility in terms of meshing and velocity sets encouraged us to search for efficient quadrature rules solving the weight function. This research led us to long-established cubature rules \cite{Stroud1973,Cools2003,Wilde2021}, i.e. multivariate quadratures, which are often used in Kalman filters, e.g., in \cite{Arasaratnam2009}. 

\section{Methodology}
\subsection{Compressible semi-Lagrangian LBM}
The compressible SLLBM uses the established lattice Boltzmann equation with the BGK collision operator \cite{Kramer2017}
\begin{equation}\label{eq:LBM}
    h_i(\mathbf{x}\!+\!\delta_t \boldsymbol{\xi}_i, t \!+\! \delta_t) = h_i(\mathbf{x},t) - \frac{1}{\tau}\!\left(h_i\!\left(\mathbf{x},t\right) - h_i^\mathrm{eq}\!\left(\mathbf{x},t\right)\right).
\end{equation}
Here, $h_i$ denotes either $f_i$ or the second distribution function $g_i$ related to the variable heat capacity ratio $\gamma$. The shifted dimensionless relaxation parameter $\tau=\lambda/\delta_t+0.5 = \mu/(P c_s \delta_t) +0.5$ depends on dynamic viscosity $\mu$, lattice speed of sound $c_s$, and pressure $P=\rho T$ with density $\rho$ and temperature $T$. The discrete equilibrium distribution function $f_i^\mathrm{eq}$ is 
\begin{equation} \label{eq:discreteEquilibrium}
f_i^{\mathrm{eq},N}(\mathbf{x},t) = w_i \sum_{n=0}^N \frac{1}{n!}\boldsymbol{a}^{(n)}_{eq}(\mathbf{x},t) :  \boldsymbol{\mathcal{H}}_i ^{(n)},
\end{equation}
and $g^\mathrm{eq} = (2C_v - D) T f_i^\mathrm{eq}$, with heat capacity at constant volume $C_v$, and number of dimensions $D$. Both $ \boldsymbol{a}^{(n)}_{eq}$ and $\boldsymbol{\mathcal{H}}_i ^{(n)} := \boldsymbol{\mathcal{H}}^{(n)}(\boldsymbol{\xi}_i)$ are listed in Appendix \ref{app:moments} and \ref{app:hermite}, respectively. To adjust the heat conductivity, a quasi-equilibrium approach \cite{Thantanapally2013} is applied to Eq. \eqref{eq:LBM}, for more details see Appendix \ref{app:quasi}.

\begin{figure}
    \centering
    \includegraphics{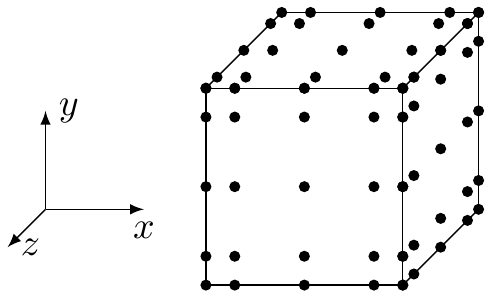}
    \caption{Support points of a three-dimensional reference cell with Gauß-Lobatto-Chebyshev points of order $p=4$. Interior and rear points are not shown.}
    \label{fig:reference_cell}
\end{figure}

Density $\rho$, momentum $\rho u$, and energy $E$ are determined by
\begin{equation} \label{eq:moments}
    \rho\! =\! \sum_{i=0}^{Q-1} f_i,\;  \rho u\! =\! \sum_{i=0}^{Q-1} f_i \boldsymbol\xi_i,\; 2 \rho E = \sum_{i=0}^{Q-1} \left(f_i |\boldsymbol\xi_i|^2 \!+ g_i \right).
\end{equation}
Appendix \ref{app:chapman} demonstrates the connection of the present SLLBM model with the macroscopic equations using a Chapman-Enskog analysis \cite{chapman1970}.
The integration along characteristics, hidden in Eq.~\eqref{eq:LBM}, incorporates a second-order temporal error, whose order can be increased by multistep schemes \cite{Wilde2019a}. In standard LBMs, the particle velocities $\boldsymbol\xi_i$ in Eq. \eqref{eq:LBM} are designed to end up on one of the neighboring nodes, and the time step size is invariably set to unity for the same purpose. By contrast, the SLLBM's particle distribution functions are still integrated along characteristics, but the departure points may reside offside the grid, i.e. they are off-lattice. To recover the off-lattice values an interpolation is needed. While several interpolation strategies are possible, we chose a cell-oriented approach, which means that once a departure point is identified, the degrees of freedom points $\hat{h}_{i \Xi j}$ in the enclosing cell are used for the interpolation \vspace{-.3cm}
\begin{equation}\label{eq:interpolation}
 h_i(\mathbf{x},t) = \sum_{j=1}^{N_{j}} \hat{h}_{i\Xi j}(t) \psi_{\Xi j}(\mathbf{x})
 \vspace{-.1cm}
\end{equation}
in cell $\Xi$ and with the basis functions $ \psi_{\Xi j}$. A three-dimensional reference cell with polynomial order $p=4$ is shown in Fig. \ref{fig:reference_cell}.
For each of the $N$ support points in the simulation, there are $Q$ particle velocities, i.e. there are $N \cdot Q$ departure points to be identified. Therefore, at the beginning of the simulation the path from each support point to the corresponding $Q$ departure points is tracked through all adjacent cells. Then a sparse matrix $\boldsymbol\Psi$ stores the interpolation coefficients $\psi_{\Xi j}$ belonging to the departure point's position; the algorithm is presented in \cite{Kramer2020}. The actual streaming step is expressed as a matrix-vector multiplication 
\begin{equation}
    \boldsymbol h_i(t+\delta_t) = \boldsymbol\Psi_i \boldsymbol h_i(t),
\end{equation}
whereas the collision step remains local.

\subsection{Cubature-based velocity sets}
The discretization of the velocity space is a key principle for any simulation with the lattice Boltzmann method. If a quadrature is to be applied, it must be suited to integrate the weight function $\mathrm{exp}{(-x^2)}$, and it has to be of ninth degree to enable compressible flow simulations \cite{Shan2006}. A prominent method to derive two- and three-dimensional velocity sets is the Gauß-product rule applied to a one-dimensional quadrature. Application to the one-dimensional degree-nine Gauß-Hermite quadrature delivers a two-dimensional D2Q25 off-lattice velocity set with 25 abscissae, which we used for previous work \cite{Wilde2019}. Due to its structure, this velocity set is infeasible for standard on-lattice streaming but perfectly suited for the SLLBM. 

\begin{figure}
\vspace{-.8cm}
    \centering
    \includegraphics[width=7.4cm]{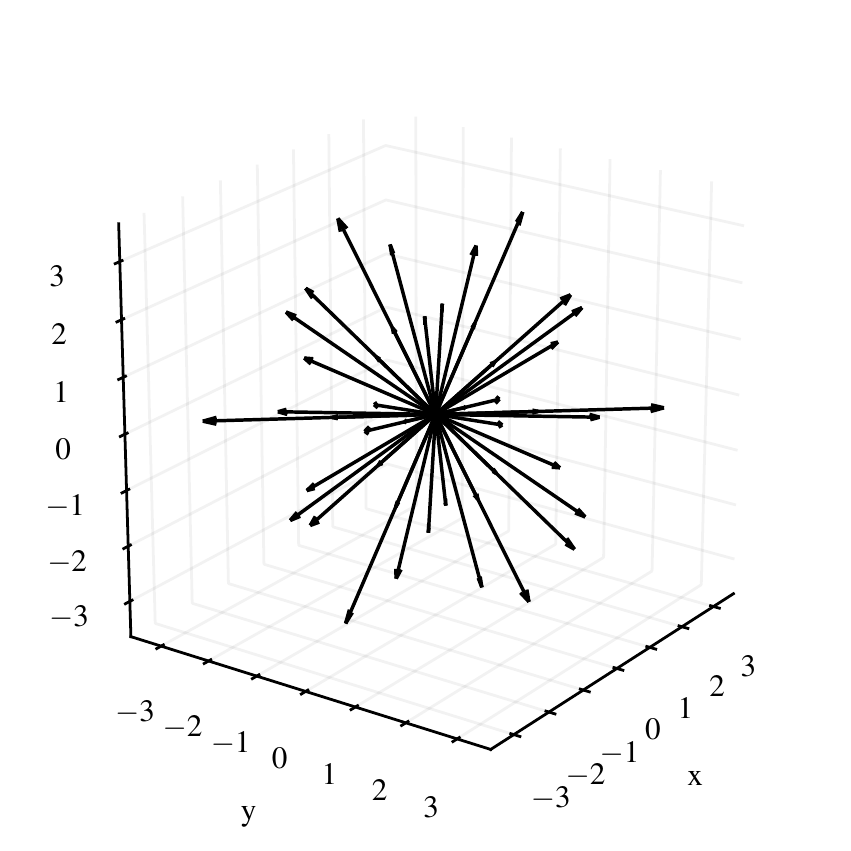}
    \vspace{-.3cm}
    \caption{Three-dimensional D3Q45 velocity set with 45 abscissae, derived by a degree-nine cubature rule originally by Konyaev \cite{konyaev1977}}
    \label{fig:D3Q45}
\end{figure}

For the simulations in this work we used velocity sets derived by cubature rules \cite{Stroud1973,Cools2003}, exhibiting the same degree of precision but consisting of fewer support points to lower computational cost. In two dimensions we employed the degree-nine D2Q19 velocity set with 19 abscissae \cite{Haegemans1977} that we have presented in recent work \cite{Wilde2021}.
In three dimensions, the Gauß-product rule led to noncompetitive 125 abscissae for a three-dimensional degree-nine D3Q125 velocity set. Therefore, we successfully identified and implemented a degree-nine D3Q45 velocity set with only 45 components, which fulfills all requirements in terms of symmetry. The D3Q45 was derived by a cubature rule after Konyaev \cite{konyaev1977} and has recently been listed in the supplemental material of a cubature article by van~Zandt  \cite{VanZandt2019}. The resulting discrete velocities are shown in Fig. \ref{fig:D3Q45}, whereas weights and abscissae of the D2Q19 and D3Q45 are listed in \cite{Wilde2021}. 

\section{Results}
We test the proposed method through three test cases. The first two test cases, a temporally underresolved Sod shock tube and a two-dimensional Riemann problem, demonstrate the effect of extraordinarily large time step sizes on the simulations. The third test case is the compressible three-dimensional Taylor-Green vortex at different Reynolds and Mach numbers to present the capability of SLLBM to simulate both turbulent and shocked three-dimensional flows.
\subsection{Temporally underresolved Sod shock tube} \label{sec:shocktube}
The Sod shock tube illustrates the large range of time step sizes accessible to the SLLBM. The domain $x\in[0,1]$ was divided into two regions at $x=0.5$ with initially
\begin{eqnarray*}
    \rho_0 = 8, u_0=0, P_0 = 10, \\
    \rho_1 = 1,u_1 = 0, P_1 = 1,
\end{eqnarray*}
and a viscosity of $\mu=7\cdot 10^{-4}$. 
\begin{figure}
    \centering
    \includegraphics{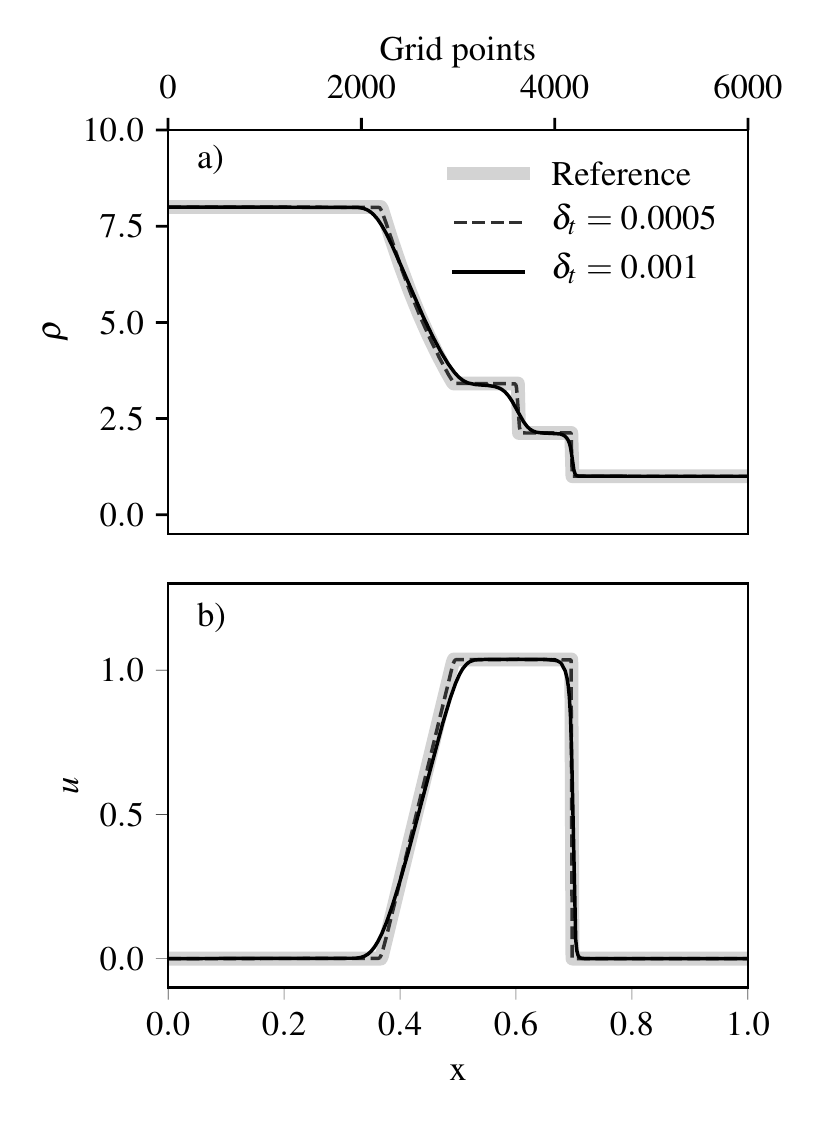}
    \vspace{-.5cm}
    \caption{Density a) and velocity b) of Sod shock tube simulation using SLLBM with 2000 cells at time $t=0.1$, which is reached after only 100 time steps ($\delta_t=0.001$), despite the fine spatial resolution of 6000 grid points. For a smaller time step size and lower viscosity the reference is perfectly matched.}
    \label{fig:shocktube}
    \vspace{-.2cm}
\end{figure}
\begin{figure}
\vspace{0.4cm}
    \centering
    \includegraphics{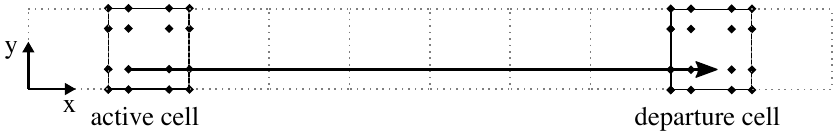}
    \vspace{-0.1cm}
    \caption{Exemplary departure point location for the Sod shock tube configuration in this article with 2000 cells. The path of the longest abscissa of the 2D velocity set D2Q19 is shown. Starting from the current cell, the abscissa's path linearly traverses six contiguous cells and locates the departure point in the seventh cell. A 3rd-order polynomial interpolation using the cell-local Gauß-Lobatto-Chebyshev support points is applied to reconstruct the distribution function value.}
    \label{fig:departure}
\end{figure}
The domain $x\in[0,1]$ was discretized using 2000 cells at third polynomial order, i.e. 6000 grid points in x direction. Despite this fine spatial resolution, which we only chose for demonstration purposes, the time step size was set to $\delta_t=0.001$, such that the solution at $\mathrm{t}=0.1$ was reached by performing only 100 time steps. 
Fig. \ref{fig:shocktube} shows that the SLLBM accurately traces the shock front despite the extremely large time step. For a smaller time step size of $\delta_t=0.0005$ and viscosity of $\mu=10^{-5}$ the simulation results perfectly matched the inviscid reference solution.
To get an impression of the time step size of other solvers, we repeated the simulations with the lower viscosity $\mu=7\cdot 10^{-4}$ using the same collision process and velocity discretization, but this time applying a spectral-element discontinuous Galerkin solver for the streaming step \cite{Min2011,Kramer2020}. This solver also features high-order solutions, but requires a dedicated time integrator and the time step size is bounded by the CFL condition \eqref{eq:CFL}. The simulation produced nearly identical results as the SLLBM with $\delta_t=0.001$ (therefore not shown), but required $\delta_t=0.000086$ to be numerically stable, i.e., 1154 time steps with an explicit exponential time integrator \cite{Uga2013} and $\delta_t=0.00005$, i.e., 2200 time steps with the more common fourth-order Runge-Kutta method. 

\begin{figure}
    \centering
    \includegraphics{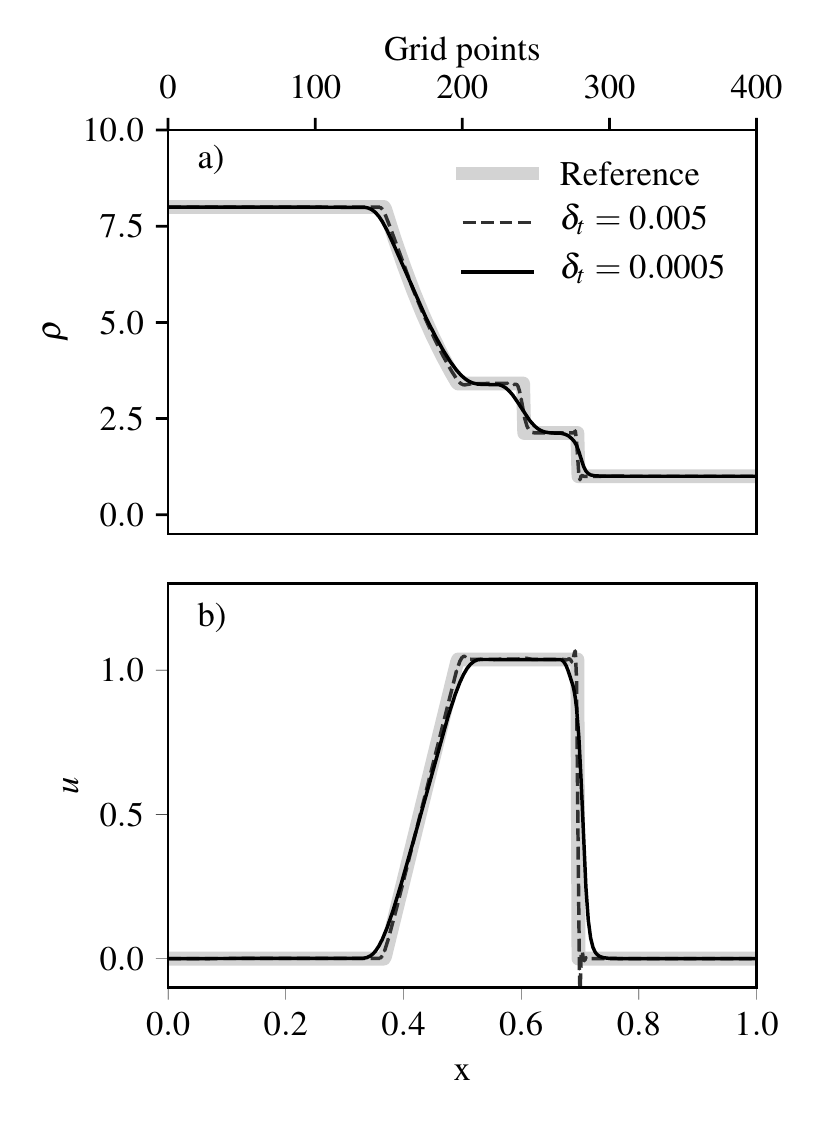}
    \caption{a) Density b) velocity of Sod shock tube simulation using SLLBM with 100 cells at time $t=0.1$, which is reached after 20 time steps ($\delta_t=0.005$), at a more reasonable resolution of 400 grid points. Simulations with smaller time step sizes and viscosity match the reference better, despite slight overshoots visible near the shock fronts.}
    \label{fig:shocktube400}
\end{figure}

As the distance of the departure points from the active nodes is proportional to the time step size (Eq. \ref{eq:LBM}), the departure points of the SLLBM were located up to seven cells away. 
The trajectory is shown in Fig. \ref{fig:departure} for an exemplary departure point. It is obvious that the CFL restriction of explicit Eulerian solvers prohibits the exchange of information crossing multiple cells. 
This property is of special interest in the case of simulations with body-fitted meshes, where the spatial extent of the smallest cells usually dictates the time step size of the whole simulation. As opposed to Eulerian solvers, the maximum stable time step size in the SLLBM is proportional only to the viscosity and not dictated by the mesh size. On top of that, when doubling the number of cells the number of SLLBM time steps can still be kept constant, whereas it inevitably doubles for the explicit  discontinuous Galerkin solver. 
Finally, Fig. \ref{fig:shocktube400} confirms that the SLLBM is also capable to stably simulate the shock tube with a lower resolution of 100 cells, polynomial order $p=4$ and 400 grid points, without additional numerical dissipation measures. Here, the time step size was set to $\delta_t=0.005$ and $\delta_t=0.0005$ with viscosities $\mu=0.002$ and $\mu=0.0002$, respectively. 

\subsection{2D Riemann study of time step size effects}
\begin{figure*}[t]
    \centering
    \includegraphics{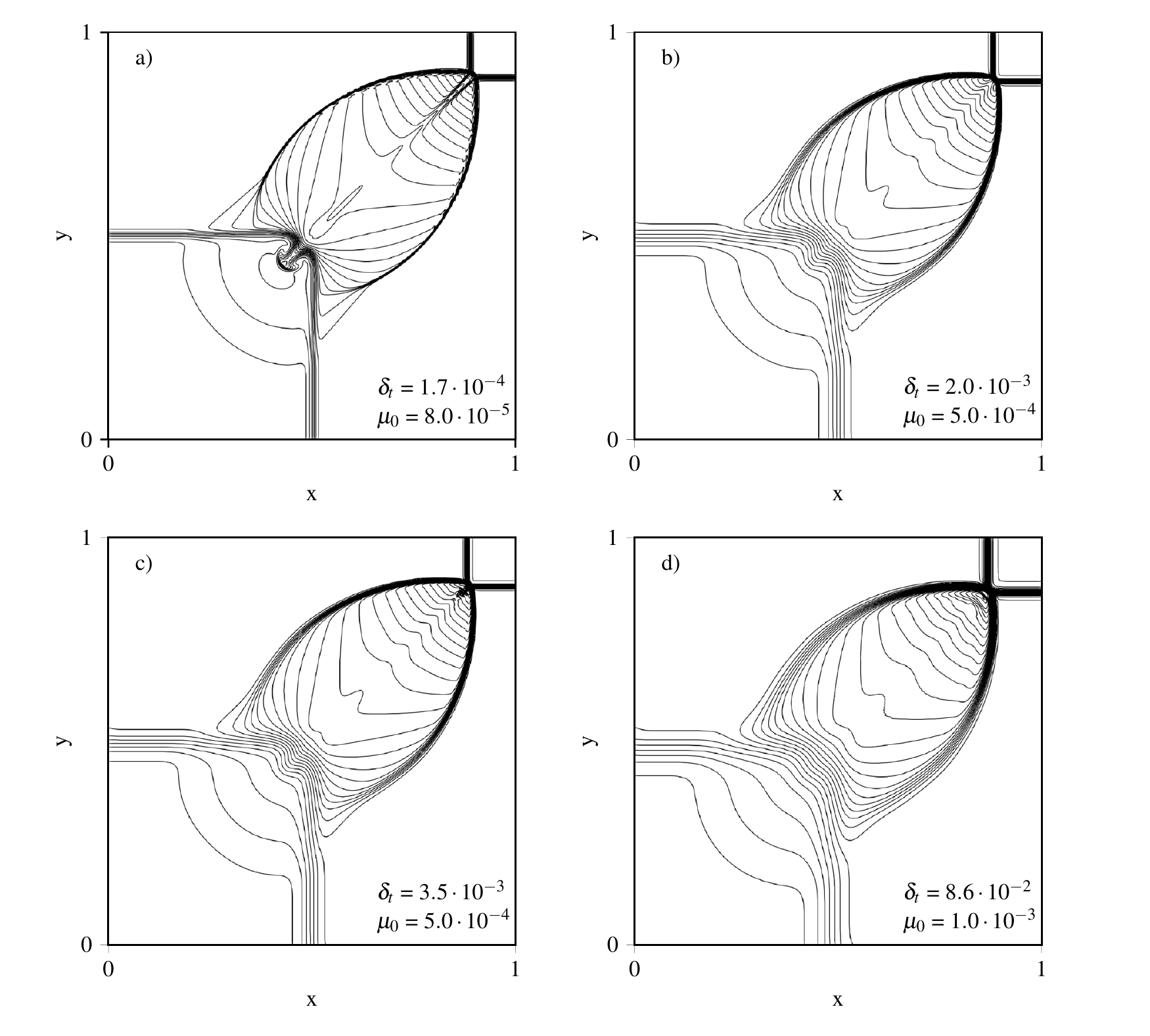}
    \caption{Density contours of the 2D Riemann simulations using $512 \times 512$ grid points.  The time step size of simulation d) is approximately 50 times larger than the time step size of simulation a), but the viscosity differs by a factor of approximately 12. As a general rule, lower viscosities demand smaller time step sizes. When choosing the time step size too large for a given viscosity, the simulations tend to become unstable near the shock fronts, as depicted in c). By contrast, simulation b) remains stable due to a 57 percent reduction of the time step size with equal prescribed viscosity. }
    \label{fig:riemann}
\end{figure*}

Case 12 of the two-dimensional Riemann problems was intensively studied by Lax \cite{Lax1998} as well as Kurganov and Tadmor \cite{Kurganov2002}. In one of our last works \cite{Wilde2019} we already showed that the SLLBM is capable to resolve the density contours of this test case with good visual agreement to the references. 
This time, similar to the shock tube test case in Section \ref{fig:shocktube}, we examined the effect of the time step size onto this test case. 
The initial conditions of the four quadrants are \cite{Guo2015}
\begin{align}
    (\rho, u_{x}, u_{y}, p) = \begin{cases} 
    (0.5313, 0, 0, 0.4), \; &x>0, \; y>0, \\
    (1, 0.7276, 0, 1), \; &x\leq0, \; y>0, \\
    (0.8, 0, 0, 1), \; &x\leq0, \; y\leq0, \\
    (1, 0, 0.7276, 1), \; &x>0, \; y\leq0.
    \end{cases}
\end{align}
The heat capacity ratio was $\gamma=1.4$, the number of cells was $N_\Xi = 128^2$ with polynomial order $p=4$, i.e., $N=512^2$ grid points. The simulation end time was $t_\mathrm{end}=0.25$.
Fig. \ref{fig:riemann} shows the density contours in the interval $\rho \in [0.412, 1.753]$, beginning with a small time step size and low viscosity in simulation a). Each of the figures also lists the time step size and viscosity. When increasing the time step size, the viscosity also needs to be increased to ensure stable simulations as depicted for simulation b). The effect of a maladjusted time step size is shown in simulation c), where oscillations occur in the top right corner near the shock fronts. Finally, simulation d) depicts the density contours for a approximately 50 times larger time step size than for case a). In this case, the shock fronts are widened, comparable to the observations in Section \ref{sec:shocktube}.

\subsection{Compressible Three-Dimensional Taylor-Green vortex}
To show that the SLLBM captures the intricate interactions between turbulent and compressible features including shocklets, the compressible three-dimensional Taylor-Green vortex flow was simulated. On the triply-periodic domain $S=[0,2\pi]^3$, the initial conditions with constant temperature initial condition (CTIC) are

\begin{align}
	u_1 (x_1, x_2, x_3, t=0) &= \sin x_1 \cos x_2 \cos x_3,\\
	u_2 (x_1, x_2, x_3, t=0) &= -\cos x_1 \sin x_2 \cos x_3 ,\\
	u_3 (x_1, x_2, x_3, t=0) &= 0,\\
	\rho (x_1, x_2, x_3, t=0) &= 1 + \frac{\mathcal{C}}{16} [\cos(2 x_1)  \\  &+\cos(2x_2)] \left[ \cos (2 x_3 + 2) \right], \\
	T(x_1, x_2, x_3, t=0) &= 1, 
\end{align}

with velocities $\mathbf{u}$, density $\rho$, Mach number $\mathrm{Ma}$, and temperature $T$. The numerator $\mathcal{C}$ differs between the cases in this work with $\mathcal{C}=1$ for $\mathrm{Re}=100$ and $\mathrm{Re}=400$ as well as $\mathcal{C}=\gamma\mathrm{Ma}^2$ for $\mathrm{Re}=1600$. The Reynolds number is defined as $Re=1/\nu_\infty$, where the subscript $\infty$ denotes the value at $T=1$. The Prandtl number is $\mathrm{Pr}=0.7$; the heat capacity ratio is $\gamma = 1.4$. The dynamic viscosity $\mu = \nu \rho $ obeys the Sutherland law

\begin{equation}
    \mu = \frac{1.4042T^{1.5}}{T+0.40417}\mu_\infty.
\end{equation}

In comparison to forced or decaying isotropic turbulence, this test case enables a deterministic initialization and thus an easier and more objective comparison. 

\subsubsection{Reynolds number $\mathrm{Re}=400$}
Peng and Yang thoroughly studied the compressible Taylor-Green vortex at Reynolds number $\mathrm{Re}=400$ \cite{Peng2018}. The original work used a compact eighth-order finite difference scheme \cite{Lele1992} to discretize the convective terms in the Navier-Stokes equation in combination with a seventh-order weighted essentially non-oscillatory (WENO) scheme. The present compressible SLLBM uses fourth-order polynomials for the equilibrium (Eq. \eqref{eq:equilibrium}) and for the interpolation (Eq. \eqref{eq:interpolation}), \emph{without utilizing additional filtering or stabilizing} techniques. Moreover, to satisfy the CFL condition, the original work applied a time step size of $\delta_t = 0.0005$, wheares the SLLBM was capable to utilize time step sizes two orders of magnitude larger: $\delta_t=0.017$ for Mach number $\mathrm{Ma}=0.5$,   $\delta_t=0.033$ for $\mathrm{Ma}=1.0$,   $\delta_t=0.018$ for $\mathrm{Ma}=1.5$, and $\delta_t=0.012$ for $\mathrm{Ma}=2.0$. The spatial resolution was $N_\mathrm{points} = 256^3$, whereas the reference operated with $512^3$ grid points.

\begin{figure}
    \centering
    \hspace{.3cm}\includegraphics[scale=1]{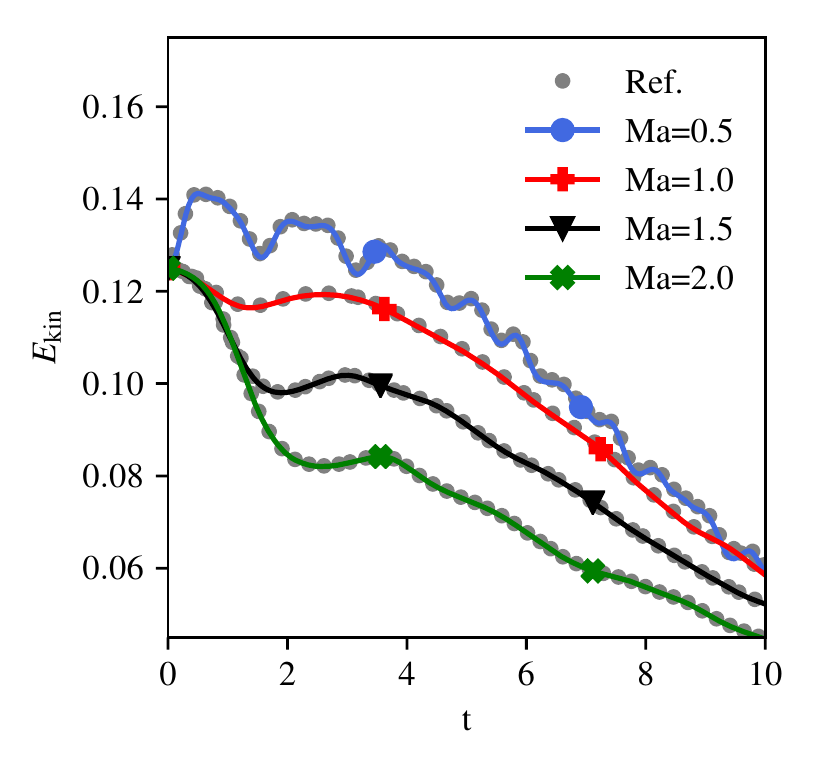}
    \vspace{-.2cm}
    \caption{Kinetic energy over time of the compressible 3D Taylor-Green vortex at Reynolds number $\mathrm{Re}=400$. Simulations by the SLLBM with the D3Q45 velocity set shown in Fig.~\ref{fig:D3Q45}. Reference from \cite{Peng2018}.}
    \label{fig:kinEnergy}
    \vspace{-.4cm}
\end{figure}

\begin{figure}
    \centering
    \includegraphics{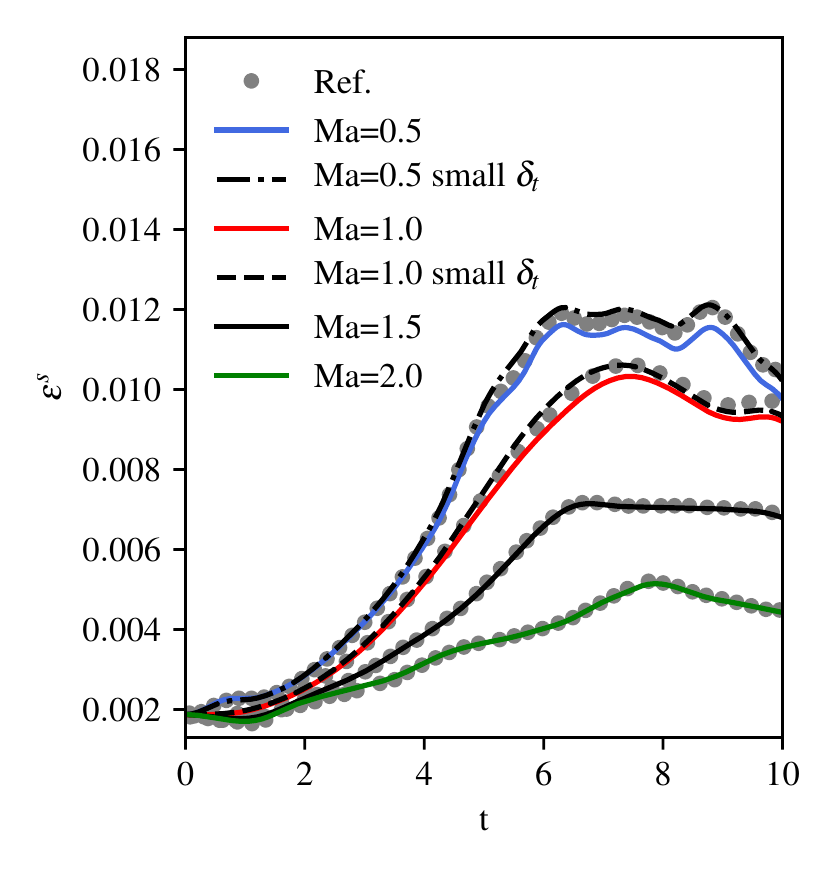}
    \caption{Solenoidal dissipation $\varepsilon^s$ of the compressible 3D Taylor-Green vortex at Reynolds number $\mathrm{Re=400}$. Small time step sizes at $\mathrm{Ma}=0.5$ and $\mathrm{Ma}=1.0$ with 5.5 times and 11.1 times smaller times step sizes, respectively. The reduction induced a slightly better agreement with the reference from \cite{Peng2018}.}
    \label{fig:400solenoidal}
\end{figure}

Fig. \ref{fig:kinEnergy} depicts the kinetic energy over time for all Mach numbers, which accurately follows the reference solution despite the coarse temporal discretization. Next, Figure \ref{fig:400solenoidal} depicts the solenoidal dissipation defined as 

\begin{equation}
    \varepsilon^s = \frac{1}{\rho_\mathrm{ref} \mathrm{Re}}\int_S \mu (\nabla \times \mathbf{u})^2 \  d^3 \mathbf{x}.
\end{equation}
 
All Mach numbers quite accurately matched the reference. The effect of changing the time step to similar levels as in the above mentioned reference can be seen for the smaller Mach numbers $\mathrm{Ma}=0.5$ and $\mathrm{Ma}=1.0$. The small deviation from the reference seen here is reduced by decreasing the time step size to similar levels, here $\delta_t=0.003$ in both cases.

\begin{figure}
    \centering
    \includegraphics{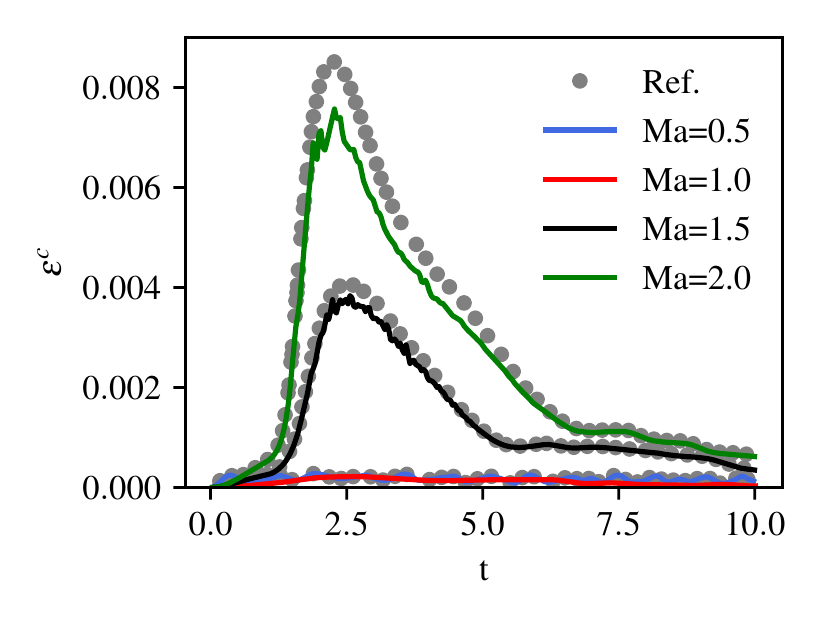}
    \caption{Dilatational dissipation $\varepsilon^c$ of the compressible 3D Taylor-Green vortex at Reynolds number $\mathrm{Re=400}$. The time step sizes are identical to the large $\delta_t$ simulations in Fig. \ref{fig:400solenoidal}.}
    \label{fig:400dilatational}
\end{figure}

The dilatational dissipation 
\begin{equation} \label{eq:dilatational}
    \varepsilon^c = \frac{4}{3\rho_\mathrm{ref} \mathrm{Re}}\int_S \mu (\nabla \cdot \mathbf{u})^2 \  d^3 \mathbf{x}
\end{equation}
 is a measure for pressure work in the simulation. Fig.~\ref{fig:400dilatational} shows that the dilatational effects are strong in the beginning at small simulation times, surmounting the solenoidal dissipation. The comparison to the reference shows a slight deviation from the reference and mild oscillations near the peak values around $t=2.5$.

\subsubsection{Reynolds number $Re=1600$}
\begin{figure}
    \centering
    \includegraphics{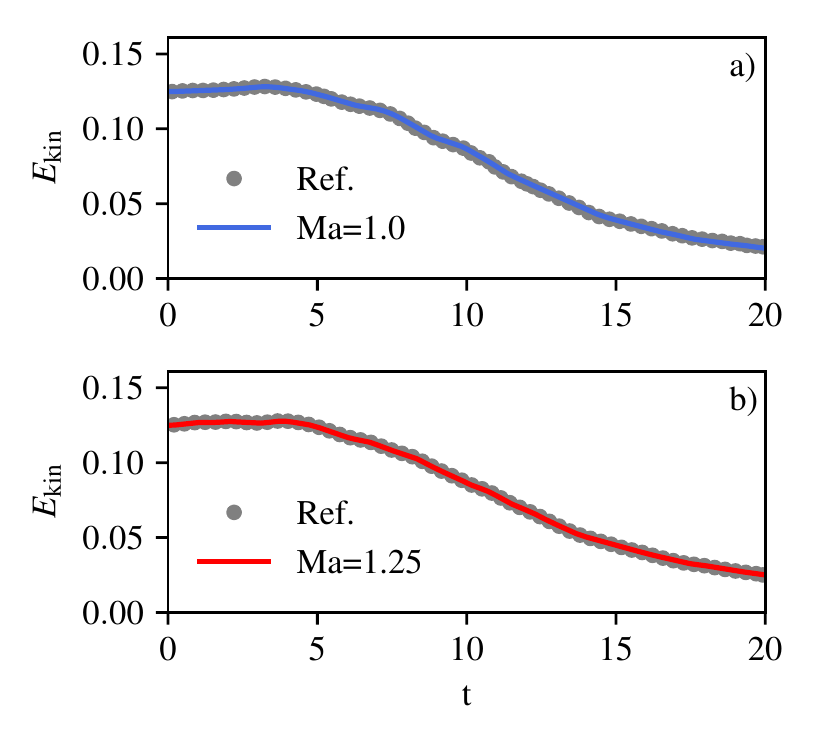}
    \caption{Kinetic energy over time of the compressible 3D Taylor-Green vortex at Reynolds number {$\mathrm{Re}=1600$} for Mach numbers a) $\mathrm{Ma}=1.0$ and b) $\mathrm{Ma}=1.25$. Reference from \cite{Lusher2021}.}
    \label{fig:1600kinEnergy}
\end{figure}

\begin{figure}
    \centering
    \includegraphics[width=0.95\linewidth]{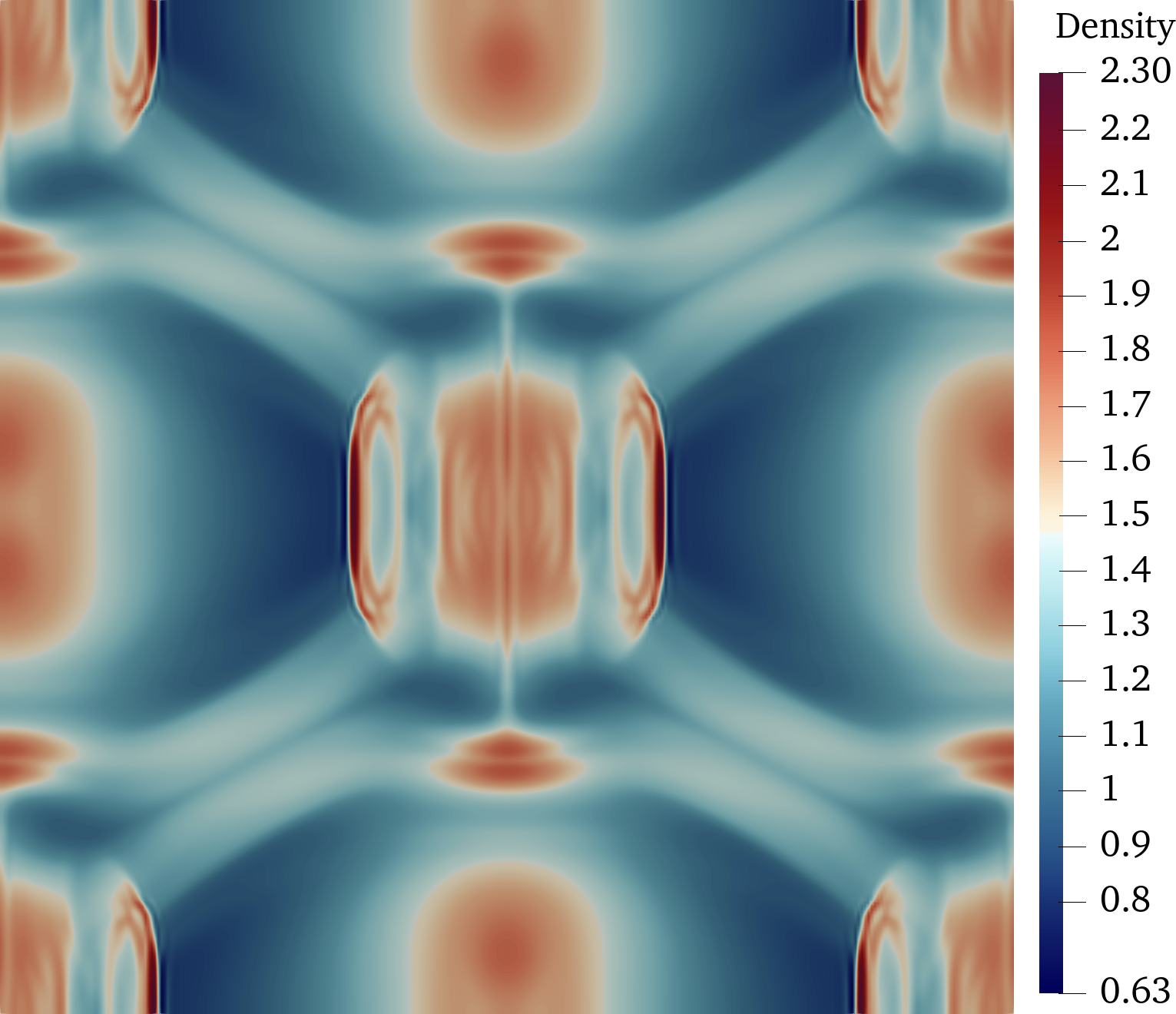}
    \caption{Density of the three-dimensional Taylor-Green vortex with $\mathrm{Re}=1600$ and Mach number $\mathrm{Ma}=1.25$ at time $t~=~3.34$. The x-z plane is shown at position $y=\pi$.}
    \label{fig:TGV-density}
\end{figure}

The Taylor-Green vortex at Reynolds number $\mathrm{Re}~=~400$ shows strong dilatational effects, but the transition to fully developed turbulence requires higher Reynolds numbers. Therefore, another recent work by Lusher and Sandham examined this test case at Reynolds number $\mathrm{Re}=1600$ up to Mach number $\mathrm{Ma}=1.25$ \cite{Lusher2021}. In their study, the authors compared high-order finite difference schemes equipped by WENO or targeted essentially non-oscillatory (TENO) schemes of different orders. 
The present work re-examines the Mach numbers $\mathrm{Ma}=1.0$ and $\mathrm{Ma}=1.25$ up to $t=20$, with time step sizes $\delta_t=0.010$, and $\delta_t=0.009$, respectively. These time step sizes were still 20 times larger than those in the reference \cite{Lusher2021}. The resolution was $N_\mathrm{points}=256^3$, whereas the reference used $512^3$ grid points. Fig. \ref{fig:1600kinEnergy} demonstrates that, just as in the case $\mathrm{Re}=400$, the kinetic energy over time is excellently reproduced for both Mach numbers.
\begin{figure}
    \centering
    \includegraphics{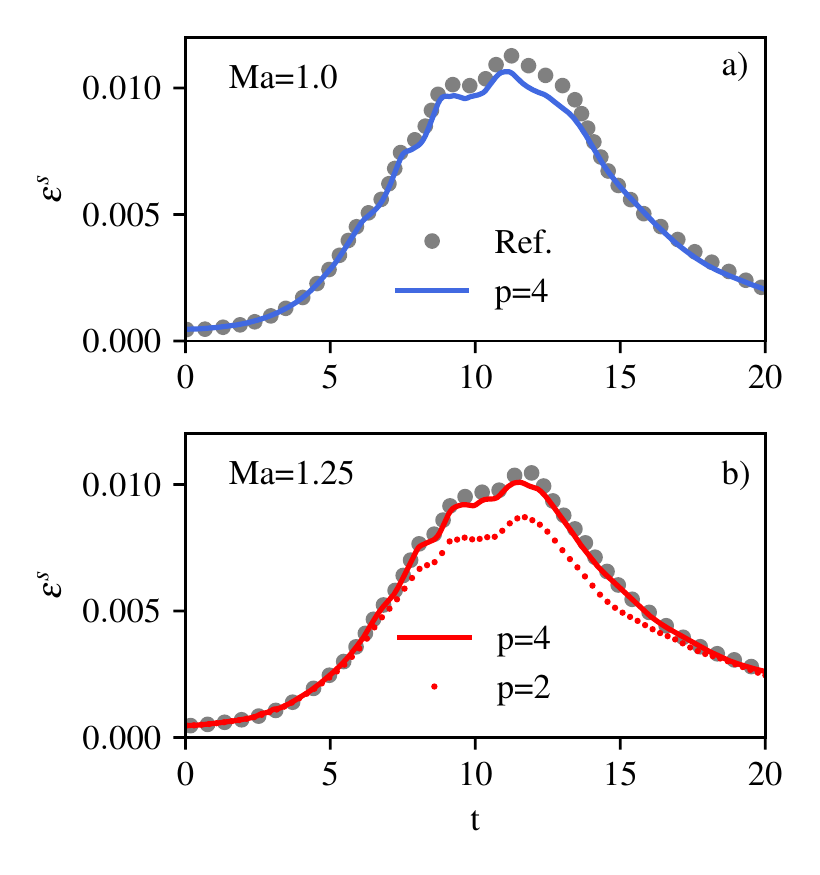}
    \caption{Solenoidal dissipation $\varepsilon^s$ of the compressible 3D Taylor-Green vortex at Reynolds number $\mathrm{Re=1600}$ for Mach numbers a) $\mathrm{Ma}=1.0$ and b) $\mathrm{Ma}=1.25$. Note the differences between second-order polynomials $p=2$ and fourth-order polynomials $p=4$.}
    \label{fig:1600solenoidal}
\end{figure}
Fig. \ref{fig:1600solenoidal} shows the solenoidal dissipation in comparison to the reference with even better results for the higher Mach number  $\mathrm{Ma}=1.25$. The good agreement confirms the little numerical dissipation introduced by the SLLBM at polynomial order $p=4$ in turbulent flows at transonic Mach numbers. By contrast, the interpolation order $p=2$ worsens the solution despite the identical resolution of $N_\mathrm{points}=256^3$.
\begin{figure}
    \centering
    \includegraphics{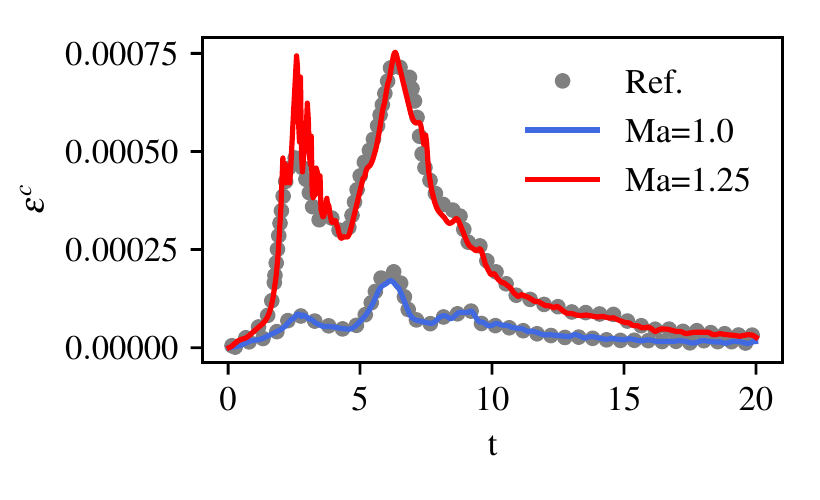}
    \caption{Dilatational dissipation $\varepsilon^c$ of the compressible 3D Taylor-Green vortex at Reynolds number $\mathrm{Re=1600}$.}
    \label{fig:1600dilatational}
\end{figure}
Next, we investigated the dilatational dissipation. During the early phase of the simulation the large vortex structures begin to entangle. This entanglement leads to strong moving shock-like structures or turbulent shocklets with strong negative dilatation \cite{Lusher2021} and local numerical oscillations of the macroscopic variables due to an underresolution of these shocklets. A slice of the density field illustrating the shocklets, indicated by large jumps in the governing variables, is shown in Fig. \ref{fig:TGV-density}. The size of the jumps agreed well with that obtained via classical Rankine-Hugoniot jump conditions. Additionally, during the early phase, the Mach numbers can be higher than the initially prescribed Mach numbers. Moreover, to compute Eq. \eqref{eq:dilatational}, we made use of the gradients of the interpolation polynomials, which---in contrast to the distribution functions---are not continuous over the cells. It is likely that this approach to compute the dilatational dissipation faces issues at strong shocklets. This explains the oscillations for $\mathrm{Ma}=1.25$ in Fig. \ref{fig:1600dilatational}, which depicts the dilatational dissipation. Despite these deviations from the reference in the beginning, the SLLBM was able to reproduce the dilatational dissipation well for the rest of the simulation, as shown in Fig. \ref{fig:1600dilatational}, without using additional stabilizing measures like filtering or artificial diffusivity. 

\subsubsection{Reynolds number $\mathrm{Re}=100$ for higher Mach numbers}
\begin{figure}
    \centering
    \includegraphics{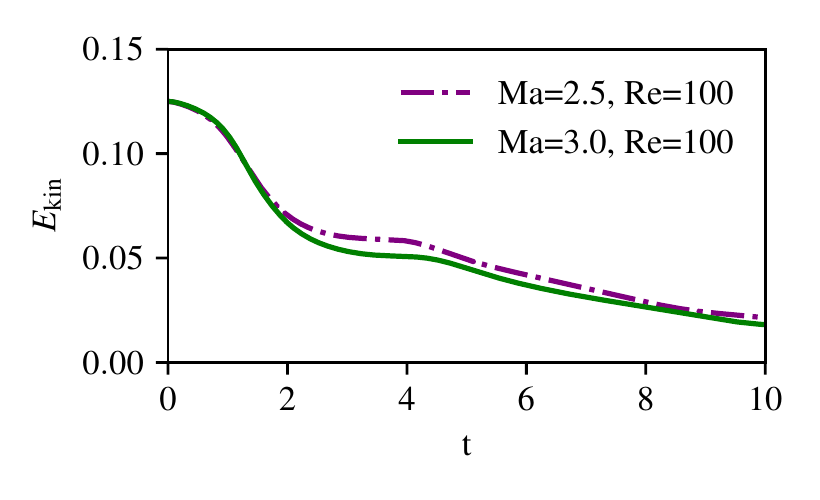}
    \caption{Kinetic energy over time of the compressible 3D Taylor-Green vortex at Reynolds number $\mathrm{Re}=100$ for Mach numbers $\mathrm{Ma}=2.5$ and $\mathrm{Ma}=3.0$.} 
    \label{fig:kinEnergyMa30}
\end{figure}

As a last test case, Fig. \ref{fig:kinEnergyMa30} shows that for $\mathrm{Re}=100$ we were able to stably simulate the given test case for Mach numbers $\mathrm{Ma}=2.5$ and $\mathrm{Ma}=3.0$. The Knudsen number  for this configuration was of order $\mathrm{Kn}=\mathcal{O}(\mathrm{Ma}/\mathrm{Re}) = \mathcal{O}(10^{-2})$. Due to this rather large Knudsen number, the SLLBM was once more able to use large times steps $\delta_t=0.015$ and $\delta_t=0.03$, respectively. Although only shown at low Reynolds numbers, these stable simulations indicate the principal capability of the SLLBM to perform stable simulations at high Mach numbers.

\section{Discussion}

As demonstrated by the numerical experiments, the SLLBM is able to simulate complex two- and three-dimensional compressible flows with shocks and shocklets. There are two main arguments in favor of the method. First, the method works with large and adjustable time step sizes. Second, cubature rules provide lean velocity sets resulting in an efficient scheme. The D3Q45 used in this work is, to the authors knowledge, the smallest known degree-nine velocity set and key to reduce errors caused by interpolation.

The SLLBM is not subject to a CFL condition, as already shown in past works \cite{Kramer2017} and as confirmed by the shock tube simulations in this work. In contrast to the usual spatial filtering, the independence of the CFL condition opens the field of temporal large eddy simulations (TLES) with fine spatial, but coarse temporal resolutions \cite{Pruett2008,Oberle2020}. Since no dedicated filter operation has been used, we classify the scheme as some sort of temporal implicit large eddy simulation (TILES). Despite the temporally coarse resolutions, most relevant flow features during the simulations were preserved, or they were recovered by scaling the time step size down: a key feature of the SLLBM that on-lattice Boltzmann schemes often miss. 

Although there is no CFL condition, the simulations are not automatically stable, because the collision step may still be prone to instabilities at small relaxation times as pointed out by various works, e.g. by Siebert et al. \cite{Siebert2008}.  That said, the stability regions of the BGK collision operator combined with the D3Q45 and fourth-order equilibria (at high Mach numbers and adjustable time step sizes) are yet unknown. Nevertheless, we observed that the SLLBM produces good results without using additional filtering or stabilizing techniques. To accomplish simulations with even higher Reynolds and Mach numbers with large time step sizes, the application of such techniques to the SLLBM might be beneficial, though.

\subsection{Comparison to other LBM solvers}
Let us now compare the SLLBM for three-dimensional compressible flows to other LBM solvers, with distinction between on-lattice solvers and off-lattice solvers. Since we discussed many aspects in recent works \cite{Wilde2019,Wilde2021}, we mainly restrict the discussion to compressible LBM solvers with applications to three-dimensional turbulent flows.

To perform a time step, off-lattice Boltzmann methods typically require a special treatment of the distribution function values such as the discretization by finite difference schemes \cite{Sun2003,Watari2006,Noah2021} or finite volume schemes \cite{Guo2015} and the application of a dedicated time integrator, e.g., a Runge-Kutta scheme. As an example for these Eulerian time integration schemes, Chen et al. presented compressible decaying three-dimensional isotropic turbulence simulations obtained by a modified discrete unified gas kinetic scheme (DUGKS), which is essentially a finite volume LBM with second-order spatial and temporal accuracy \cite{Chen2020}. Also, the authors made use of a high-order Gaussian quadrature to discretize the velocity space, but they relied on a decisively larger D3Q77 velocity set with identical quadrature degree as the D3Q45. Like all explicit Eulerian time integration schemes, DUGKS is subject to the CFL condition \eqref{eq:CFL}. 

A second category of off-lattice schemes are interpolation-based or semi-Lagrangian implementations. Pavlo et al. pioneered in using second-order interpolations to incorporate high-order velocity sets for simulations of thermal flows \cite{Pavlo1998,Pavlo2002}. Renowned semi-Lagrangian implementations are the Particles on Demand method \cite{Dorschner2018}, which uses dynamically shifted velocity sets to reach high Mach numbers or the SLLBM for compressible flows on unstructured meshes by Saadat et al. \cite{Saadat2020}. Unlike the present method, the authors of the latter approach used a D2Q9 velocity set and computed correction terms by exploiting the gradients of the distribution functions, which practically come at low costs when using interpolation polynomials. To the best of our knowledge, these methods have not been applied to complex three-dimensional flows, except for a spherical Riemann problem presented by Zakirov et al. using a D3Q125 velocity set \cite{Zakirov2019}. Additionally, one decisive feature of the present method is the organization of support points by cells, enabling three-dimensional high-order solutions.

On-lattice Boltzmann solvers have the Lagrangian integration along characteristics in common with interpolation-based schemes, albeit constrained by a strong coupling of space and time discretization. This explains, the rather large time step sizes of on-lattice Boltzmann methods with increasing Mach number \cite{Frapolli2017, Wilde2021}. On-lattice Boltzmann methods generally exhibit second-order accuracy in space and time, but they are not as fiercely concerned by numerical diffusion as low-order off-lattice Boltzmann methods. In general, the computational costs of on-lattice Boltzmann methods' streaming step are low. On the downside, they often suffer from large velocity sets. For example, Frapolli et al. \cite{Frapolli2015,Frapolli2016a,Frapolli2020} were able to simulate various three-dimensional compressible flows, including isotropic decaying turbulence simulations or the flow around an Onera M6 wing, by the entropic lattice Boltzmann method. In many ways, their works were groundbreaking for the development of compressible LBMs. However, the authors used a velocity set with 343 discrete velocities and the weights were temperature-dependent, which limits the temperature range of the method. In addition, in many on-lattice Boltzmann frameworks the time step size is not adjustable due to the linking of spatial and temporal discretization \cite{Frapolli2017}. An exception to this are hybrid lattice Boltzmann methods, which solve the temperature field separately \cite{Nie2009,Feng2016,Feng2019,Renard2021}. In this case, by adapting the speed of sound, the time step size can be varied. By using numerical equilibria instead of the polynomial equilibrium in this work, Latt et al. were able to simulate a three-dimensional flow around a sphere \cite{Latt2020} by using only 39 discrete velocities, as similarly proposed by Frapolli \cite{Frapolli2017}. Simulations of complex turbulent flows by this method are not available, yet. Recently, Saadat et al. \cite{Saadat2021} proposed an on-lattice Boltzmann model with a regular D3Q27 velocity set to perform three-dimensional decaying compressible isotropic turbulence simulations. Despite the low degree of the velocity set, this model proved capable to simulate shocked flows up to Mach numbers of $\mathrm{Ma}~1.5$. This is achieved by correcting the error-prone high-order moments by expressions obtained from applying finite differences to the macroscopic variables. 

In summary, the number of compressible LBM solvers for three-dimensional flows is still limited for both on-lattice and off-lattice Boltzmann schemes, which indicates the value of the presented SLLBM framework to set a pattern for future research.

\subsection{Numerical efficiency}

The numerical efficiency of the SLLBM depends on the implementation of the collision and the streaming step. The former requires at each support point the calculation of the discrete equilibrium function values given by Eq. \eqref{eq:discreteEquilibrium}. Once the density, velocities, and temperature are gained from the distribution function values, the determination of the equilibrium, however, is well parallelizable. Note, that the tensors in Eq. \eqref{eq:discreteEquilibrium} are symmetric, so that many entries need to be computed only once per time step and support point, e.g., $a_{xxxy,\mathrm{eq}}$ is equal to $a_{yxxx,\mathrm{eq}}$ and all other index permutations. In addition, the entries of the Hermite tensors $\mathcal{H}_i$ are constant for a given velocity set and only need to be determined once for the whole simulation. 

Another decisive factor of the simulation performance is the streaming step, which requires a ``triquartic'' interpolation, i.e., interpolation polynomials of order $p=4$ in three dimensions. The reference cell is shown in Fig. \ref{fig:reference_cell}. To interpolate one distribution function value at a given departure point, all $(p+1)^D=125$ support points of the cell are processed with the interpolation coefficients that are stored in the $Q$ matrices $\boldsymbol\Psi_i$. These matrices are the cornerstone for the whole simulation. When accounting to equally shaped cells, like in all simulations in the present work, the size of the matrices can be significantly reduced, since the matrices are identical over nearly all cells (boundary cells excepted). However, for irregular or distorted three-dimensional meshes, the size of the matrices grows quickly, rendering the streaming step memory-bound. Therefore, matrix-free implementations \cite{Kronbichler2019} appear to be an attractive extension for SLLBM implementations. In matrix-free versions the interpolation coefficients will be computed afresh in each time step, which is potentially faster for today's high-performance computing clusters. 

At first glance, the second distribution function appears to be a severe limitation of the method. On closer inspection, however, it turns out that the overhead is kept in reasonable limits since the computation-intensive equilibrium distribution $g_i^\mathrm{eq}$ linearly depends on $f_i^\mathrm{eq}$ and the interpolation coefficients can be used for both $f_i$ and $g_i$. Still, all aforementioned issues pay off in light of the large time step sizes of semi-Lagrangian implementations. A major way to further reduce the computational cost is the reduction of discrete velocities in the Gaussian quadrature. The research for even compacter cubature rules is still ongoing \cite{Wilde2021}, so that future degree-nine velocity sets will possibly be even more efficient.

\begin{figure}
    \centering
    \includegraphics[width=1.0\linewidth]{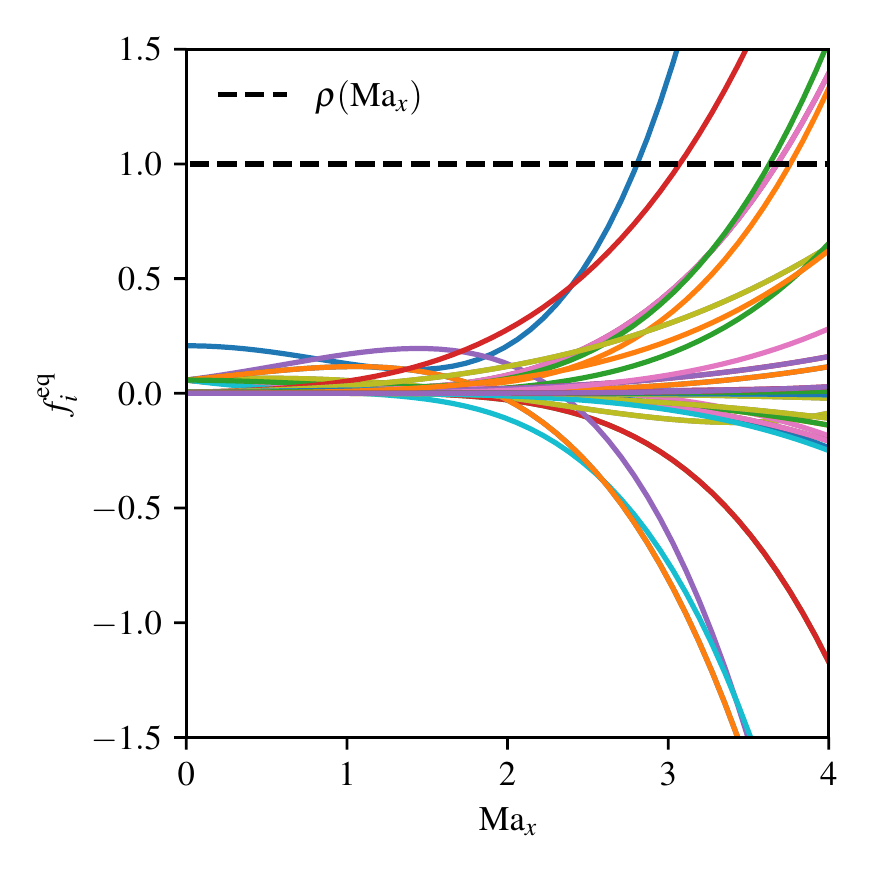}
    \caption{All 45 equilibrium distribution functions $f^\mathrm{eq}_i$ of the D3Q45 velocity set over Mach numbers ranging from 0 to 4 in x-direction. Density $\rho$ and relative temperature $\theta$ were constantly set to unity; Mach numbers in both other directions were set to zero.}
    \label{fig:equilibrium}
\end{figure}

One last remark regarding the equilibrium distribution function determined via Eq. \eqref{eq:discreteEquilibrium}. In the past the Hermite expansion-based equilibrium was deemed causing instabilities due to negative distribution function values at high Mach numbers \cite{Frapolli2017}. Fig. \ref{fig:equilibrium} depicts the equilibrium function values $f^\mathrm{eq}_i$ over the Mach number $\mathrm{Ma}_x$ in x-direction. The figure manifests that most of the discrete velocities significantly diverge for Mach numbers larger than $\mathrm{Ma}>2.0$ towards either $\infty$ or $-\infty$. However, our simulations showed that simulations even up to $\mathrm{Ma}=3.0$ remained stable. This observation indicates that negative distribution function values are not per se a source of instabilities, since the values' sole role is to encode the moments of different orders. For the first moments this encoding can also be seen in Fig. \ref{fig:equilibrium}: despite the increasing first-order moment of the momentum, the ``zeroth--order'' moment of the density remains $\rho=1.0$. To ensure this, negative values are a consequence at high Mach numbers due to the discretization of the velocity space via Gaussian quadratures.

\section{Conclusion} The SLLBM for three-dimensional compressible flows is a viable alternative to other solvers. The SLLBM allows very large time steps sizes  not  restricted by the customary CFL condition. Although the presented SLLBM is a fourth-order spatial method and accurately captures shocks as well as turbulence, no stabilization or filtering were required for the presented test cases. Due to these unique features, the cubature-based fully explicit SLLBM enables researchers to perform compressible turbulence simulations, in which the admissible time step sizes are decoupled from the spatial discretization, opening a new field of affordable simulations for compressible turbulent flows.
\begin{acknowledgments}
This work was supported by the German Ministry of Education and Research and the Ministry for Culture and Science North Rhine-Westfalia (research grant 13FH156IN6). D.W. is supported by German Research Foundation (DFG) project FO 674/17-1. \\
D.W. and A.K. contributed equally to this work.
\end{acknowledgments}

\appendix

\section{Equilibrium moments}\label{app:moments}
The equilibrium moments up to fourth order read
\begin{subequations}
\label{moments}
\begin{eqnarray}
    a^{(0)}_{eq} &=& \rho \\
    a^{(1)}_{\alpha,eq} &=& \rho u_\alpha \\ 
    a^{(2)}_{\alpha \beta,eq} &=& \Pi_{\alpha \beta}^\mathrm{eq} = \rho (u_\alpha u_\beta + T_0(\theta -1 )\delta_{\alpha\beta}) \label{moment:pi}\\
    a^{(3)}_{\alpha \beta \gamma,eq} &=& \mathcal{Q}_{\alpha \beta \gamma}^\mathrm{eq} = \rho \left[u_\alpha u_\beta u_\gamma +  T_0(\theta -1 )(\delta_{\alpha\beta}u_\gamma  \right.\nonumber \\
     & & \left.\hspace{1.2cm} + \delta_{\alpha\gamma}u_\beta +\delta_{\beta\gamma}u_\alpha)\right] \label{moment:Q}\\
    a^{(4)}_{\alpha \beta \gamma \delta,eq} &=& \mathcal{R}_{\alpha \beta \gamma \delta}^\mathrm{eq} \nonumber\\
    &=& \rho[ u_\alpha u_\beta u_\gamma u_\delta   
     + T_0(\theta-1)((\delta_{\alpha\beta}\delta_{\gamma\delta} \nonumber\\
     & &+ \delta_{\alpha\gamma}\delta_{\beta\delta}+ \delta_{\alpha\delta}\delta_{\beta\gamma} )T_0(\theta-1)  \nonumber\\ 
   & & + \delta_{\alpha\beta} u_{\gamma}u_{\delta} + \delta_{\alpha\gamma} u_{\beta}u_{\delta} + \delta_{\alpha\delta} u_{\beta}u_{\gamma} +\nonumber\\
   & & \delta_{\beta\gamma} u_{\alpha}u_{\delta} + \delta_{\beta\delta} u_{\alpha}u_{\gamma} + \delta_{\gamma\delta} u_{\alpha}u_{\beta}  ) ], \label{moment:R} 
\vspace{.3cm}
\end{eqnarray}
\end{subequations}
where $\theta = T / T_0$ is the relative temperature.

\section{Hermite tensors} \label{app:hermite}
The scaled Hermite tensors with $\hat{\xi_i} = \xi_i/c_s$ up to fourth order read 

\begin{align*}
    {\mathcal{H}_i}^{(0)}&= 1 \\
    {\mathcal{H}_{i\alpha}}^{(1)}&=
                   \frac{{{\hat{\xi}_{i\alpha}}}}{c_s}  \\
   {\mathcal{H}_{i\alpha \beta}}^{(2)}&= \frac{{{{\hat{\xi}_{i\alpha}}{\hat{\xi}_{i\beta}}}}-\delta_{\alpha\beta}}{c_s^2} \\
    {\mathcal{H}_{i\alpha \beta \gamma}}^{(3)}&= \frac{{\hat{\xi}_{i\alpha}}{\hat{\xi}_{i\beta}}{\hat{\xi}_{i\gamma}}-({\hat{\xi}_{i\alpha}\delta_{\beta\gamma}}+{\hat{\xi}_{i\beta}\delta_{\alpha\gamma}}+{\hat{\xi}_{i\gamma}\delta_{\alpha\beta}})}{c_s^3} \\ 
    {\mathcal{H}_{i\alpha \beta \gamma \delta}}^{(4)}&= \frac{{{\hat{\xi}_{i\alpha}}{\hat{\xi}_{i\beta}}{\hat{\xi}_{i\gamma}}{\hat{\xi}_{i\delta}}}-\mathcal{T}_i+(\delta_{\alpha\beta}\delta_{\gamma\delta}+\delta_{\alpha\gamma}\delta_{\beta\delta}+\delta_{\alpha\delta}\delta_{\beta\gamma})}{c_s^4},
\end{align*}

with

\begin{equation*}
\begin{split}
            \mathcal{T}_i = \hat{\xi}_{i\alpha}\hat{\xi}_{i\beta}\delta_{\gamma\delta} + \hat{\xi}_{i\alpha}\hat{\xi}_{i\gamma}\delta_{\beta\delta} + \hat{\xi}_{i\alpha}\hat{\xi}_{i\delta}\delta_{\beta\gamma} + \\ \hat{\xi}_{i\beta}\hat{\xi}_{i\gamma}\delta_{\alpha\delta} + \hat{\xi}_{i\beta}\hat{\xi}_{i\delta}\delta_{\alpha\gamma}+ \hat{\xi}_{i\gamma}\hat{\xi}_{i\delta}\delta_{\alpha\beta}.
            \end{split}
\end{equation*}

\section{Chapman-Enskog analysis} \label{app:chapman}
This section shows the approximation of the compressible Navier-Stokes equations, when applying a Chapman-Enskog analysis to the SLLBM model. To that end, a second-order multiscale expansion is applied to the temporal derivatives
\begin{align}
    \partial_\alpha &= \epsilon \partial_\alpha, \\
    \partial_t &= \epsilon \delta_t^{(1)} + \epsilon^2 \delta_t^{(2)},
\end{align}
where $\epsilon$ is a smallness parameter usually identified as the Knudsen number \cite{chapman1970} and the discrete distribution functions $h_i$
\begin{equation} \label{eq:DistributionExpansion}
    h_i = h_i^{(0)}+\epsilon h_i^{(1)} + \epsilon^2 h_i^{(2)}.
\end{equation} 
These expansion terms are applied to a second-order Taylor expansion of Eq. \eqref{eq:LBM}
\begin{equation} \label{eq:TaylorExpansion}
    (\epsilon D_i  + \frac{\epsilon^2}{2} D_i D_i) h_i = - \frac{1}{\tau}\left( h_i - h_i^\mathrm{eq}\right),
\end{equation}
with the material derivative \cite{Kullmer2019}
\begin{equation}
    \epsilon D_i = \epsilon D_i^{(1)} + \epsilon^2 D_i^{(2)} = \epsilon \partial_t^{(1)} +  \epsilon \, \xi_{i\alpha}\partial_\alpha + \epsilon^2 \partial_t^{(2)} ,
\end{equation}

Next, Eq. \eqref{eq:DistributionExpansion} is applied to \eqref{eq:TaylorExpansion} and the terms of same order are collected. The zeroth-order terms of order $\epsilon^0$ are
\begin{equation}
    0 = -\frac{1}{\tau} \left( h^{(0)} - h^\mathrm{eq} \right),
\end{equation}
with the trivial relation
\begin{equation}
    h_i^{(0)} = h_i^\mathrm{eq}.
\end{equation}
When applying the moment relations of Eq. \eqref{eq:moments} to the last Equation, we need to differ between the moments of $f_i$ and $g_i$, i.e,
\begin{equation}
\sum_i f_i^{(0)} \{1,\xi_{i\alpha}\} = \sum_i f_i^\mathrm{eq}\{1,\xi_{i\alpha}\}
\end{equation}
and
\begin{equation}
    \sum_i \left( f_i^{(0)} |\boldsymbol{\xi}_{i}|^2 + g_i^{(0)} \right) = \sum_i \left( f_i^\mathrm{eq} |\boldsymbol{\xi}_{i}|^2 + g_i^\mathrm{eq} \right),
\end{equation}
showing mass, momentum and energy conservation during the collision step, which also implies that
\begin{align}
    \sum_i f_i^{(1,2,\ldots)} \{1,\xi_{i\alpha}\} = 0,  \\ \sum_i\left( f_i^{(1,2,\ldots)} |\boldsymbol{\xi}_{i}|^2 + g_i^{(1,2,\ldots)} \right) =0 
\end{align}
As a second step, the order $\epsilon^1$ terms are collected
\begin{equation}\label{eq:CE_1stOrder}
    D_i^{(1)} h_i^{(0)} = -\frac{1}{\tau} h_i^{(1)}.
\end{equation}

This time, the relations of Eq. \eqref{eq:moments} yield for the zeroth moment
\begin{equation} \label{eq:euler_mass}
    D_i^{(1)} \rho = \partial_t^{(1)} \rho + \partial_\alpha(\rho u_\alpha) =0,
    \end{equation}
for the first moments with the help of Eq. \eqref{moment:pi} and $P = \rho(T-T_0)$
\begin{align}
    &\partial_t^{(1)} (\rho  u_\alpha) + \partial_\beta \Pi^{(0)}_{\alpha \beta} = 0 \nonumber \\ 
  &\partial_t^{(1)} (\rho  u_\alpha) + \partial_\beta (\rho u_\alpha u_\beta) + \partial_\alpha (P) \delta_{\alpha \beta} = 0,  \label{eq:euler_momentum}
\end{align}
and for the total energy  
\begin{equation}
   \partial^{(1)}_t (\rho E) + \partial_\alpha (u_\alpha \rho E) = -P \partial_\alpha u_\alpha  \label{eq:euler_energy}
\end{equation}
or
\begin{equation}\label{eq:euler_temperature}
   \partial^{(1)}_t T = -u_\alpha  \partial_\alpha T - \frac{P}{\rho C_v} \partial u_\alpha
\end{equation}
for the temperature.

These are the Euler equations; to derive the Navier-Stokes equations, the terms of order $\epsilon^2$ also need to be gathered. This leads to 
\begin{equation}
    D_i^{(2)} h_i^{(0)} + D_i^{(1)} h_i^{(1)} +  \frac{1}{2} D_i^{(1)}D_i^{(1)} h_i^{(0)} = -\frac{1}{\tau} h_i^{(2)}
\end{equation}
Eq. \eqref{eq:CE_1stOrder} simplifies the last Equation's derivatives 
\begin{equation} \label{eq:CE_2ndOrder}
    D_i^{(2)} h_i^{(0)} + \left(1-  \frac{1}{2\tau} \right) D_i^{(1)} h_i^{(1)} = -\frac{1}{\tau} h_i^{(2)}.
\end{equation}
Due to vanishing terms, the zeroth moment of Eq. \eqref{eq:CE_2ndOrder} is
\begin{equation}
    \partial_t^{(2)} \rho = 0.
\end{equation}
Then, the slightly more complicated part begins with the first moments of \eqref{eq:CE_2ndOrder}
\begin{equation} \label{eq:expansion2_order2}
\partial_t^{(2)} (\rho u_\alpha) =  -\left(1-\frac{1}{2\tau} \right) \partial_\beta \Pi_{\alpha\beta}^{(1)}.
\end{equation}
To express the nonequilibrium moment $\Pi_{\alpha\beta}^{(1)}$ by equilibrium moments, we derive the second order moment of \eqref{eq:CE_1stOrder}, which is
\begin{equation}
    \partial_t^{(1)} \Pi_{\alpha\beta}^{(0)} + \partial_\gamma \mathcal{Q}_{\alpha \beta \gamma}^{(0)} = -\frac{1}{\tau}\Pi_{\alpha\beta}^{(1)}
\end{equation}
and then complements Eq. \eqref{eq:expansion2_order2}
\begin{equation}\label{eq:pi_2nd_order}
    \partial_t^{(2)} (\rho u_\alpha) =  \left(\tau-\frac{1}{2} \right) \partial_\beta \left(\partial^{(1)}_t \Pi_{\alpha\beta}^{(0)} + \partial_\gamma \mathcal{Q}_{\alpha \beta \gamma}^{(0)} \right).
\end{equation}
The equilibrium moments can be explicitly specified by Equations \eqref{moment:pi} and \eqref{moment:Q}. By applying the product rule and by using the Euler equations for mass \eqref{eq:euler_mass} and momentum \eqref{eq:euler_momentum} and temperature \eqref{eq:euler_temperature} to replace the time derivatives, one obtains

\begin{align}
    \partial^{(1)}_t \Pi_{\alpha\beta}^{(0)} = &- \partial_\gamma (\rho u_\alpha u_\beta u_\gamma) - \frac{P}{C_v}\partial_\gamma u_\gamma \delta_{\alpha \beta} \nonumber \\
    & -\partial_\gamma (P u_\gamma)\delta_{\alpha \beta} - u_\alpha \partial_\beta P - u_\beta \partial_\alpha P \label{eq:Pi_ab}
\end{align}
and
\begin{align}
    \partial_\gamma \mathcal{Q}_{\alpha \beta \gamma}^{(0)} = &\partial_\gamma (\rho u_\alpha u_\beta u_\gamma) + P \partial_\beta u_\alpha  + P \partial_\alpha u_\beta  \nonumber \\
    &+ \partial_\gamma (P u_\gamma) \delta_{\alpha \beta} + u_\alpha \partial_\beta P + u_\beta \partial_\alpha P.  \label{eq:Q_abc}
\end{align}
This turns Eq. \eqref{eq:pi_2nd_order} into 
\begin{multline} \label{eq:rhou2}
   \partial_t^{(2)} (\rho u_\alpha) = \\ \left(\tau-\frac{1}{2} \right)\partial_\beta   \left[P (\partial_\alpha u_\beta + \partial_\beta u_\alpha) - \frac{P}{C_v} \partial_\gamma u_\gamma \delta_{\alpha \beta}  \right].
\end{multline}
   As a last step, the total energy is determined
  \begin{equation}\label{eq:totalEnergy2}
      \partial_t^{(2)} (\rho E) = - \left(1-\frac{1}{2\tau}\right) \partial_\beta q_\beta^{(1)},
  \end{equation}
  where 
  \begin{equation}
      q_\beta^{(1)} = \sum_i \frac{1}{2}|\boldsymbol{\xi}_{i}|^2 \xi_{i\beta} f_i^{(1)}
  \end{equation} 
  is a contracted variant of $\mathcal{Q}_{\alpha \beta \gamma}^{(1)}$ detailed in Eq. \eqref{moment:Q}.
  Similar to Eq. \eqref{eq:expansion2_order2}, this vector can be expressed by equilibrium counterparts
  \begin{equation}
      q_\beta^{(1)} = -\tau(\partial_t^{(0)}q_\beta^{(0)} + \partial_\gamma r_{\beta\gamma}^{(0)}).
  \end{equation}
  with 
  \begin{equation}
      r_{\beta\gamma}^{(0)} = \sum_i \frac{1}{2}|\boldsymbol{\xi}_{i}|^2 \xi_{i\beta}  \xi_{i\gamma} f_i^{(0)}.
  \end{equation}
Resembling $q_\beta$, the tensor $r_{\beta\gamma}$ is the contracted variant of $\mathcal{R}_{\alpha \beta \gamma \delta}$ detailed in Eq. \eqref{moment:R}.
Again, by a number of replacements, one obtains \cite{Coreixas2018}  
\begin{multline}
    q_\beta^{(1)} = -\tau P \bigl[(1+C_v)  \partial_\beta T \\+ \partial_\beta \bigl( u_\gamma[\partial_\beta u_\gamma + \partial_\gamma u_\beta]-u_\gamma \frac{\partial_\delta u_\delta}{C_v}\delta_{\beta\gamma} \bigr) \bigr].
\end{multline}
This term now complements Eq. \eqref{eq:totalEnergy2}. Finally, by summing up all contributions of orders $\epsilon^0$, $\epsilon^1$, and $\epsilon^2$, the compressible Navier-Stokes equations are derived
\begin{align}
    &\partial_t \rho + \partial_\alpha(\rho u_\alpha) = 0, \\
    &\partial_t(\rho u_\alpha) + \partial_\beta (\rho u_\alpha u_\beta + P\delta_{\alpha \beta}) = \partial_\beta (\sigma_{\alpha\beta}),\label{eq:NS_momentum}\\
    &\partial_t(\rho E) + \partial(\rho E u_\beta) = \partial_\beta (\lambda \partial_\beta T) + \partial_\beta (u_\gamma \sigma_{\beta\gamma}),
\end{align}
with dynamic viscosity $\mu = \tau P$, thermal conductivity $\lambda = \tau P (C_v + 1) = \tau P C_p$, since the heat capacity at constant pressure is defined as $C_p = C_v + 1 $. The stress tensor denotes
\begin{equation}\label{eq:sigma}
    \sigma_{\alpha \beta} = \mu \bigl((\partial_\alpha u_\beta + \partial_\beta u_\alpha) -P\delta_{\alpha \beta} \bigr) - \frac{\mu}{C_v}\partial_\gamma u_\gamma \delta_{\alpha \beta} .
\end{equation}
Since the dynamic viscosity depends on the local pressure $P$, the relaxation parameter has to be pressure-dependent, i.e., $\tau=\mu/(P c_s \delta_t)+0.5$. From Eq. \eqref{eq:sigma} the bulk viscosity, with respect to the notation in Eq. \eqref{eq:NS_momentum}), can be identified as $\mu_b~=~\mu / C_v$. 

Note that the presented derivation links the thermal conductivity to the dynamic viscosity. In our approach, we used the quasi-equilibrium approach to adjust the Prandtl number $\mathrm{Pr}=(\mu C_p) / \lambda$. For the corresponding extension of the Chapman-Enskog analysis, we refer to \cite{Frapolli2016}. 

\section{Quasi-equilibrium approach for variable Prandtl number} \label{app:quasi}

To obtain a variable Prandtl number $\mathrm{Pr}$, the following equation replaces Eq. \eqref{eq:LBM}
\begin{multline}
    	h_i(\mathbf{x}\!+\!\delta_t \boldsymbol{\xi}_i, t \!+\! \delta_t) = 
		h_{i}(x, t) \\ - \frac{1}{\tau} 
		\left[ h_{i}(x, t)- h_{i}^{\mathrm{eq}}(x \xi_{i}, t) \right] +  \left(\frac{1}{\tau}-\frac{1}{\tau_\mathrm{Pr}}\right)h_i^*(x , t),
\end{multline}

with $\tau_\mathrm{Pr} = (\tau -0.5)/\mathrm{Pr} + 0.5$. The quasi-nonequilibrium $h_i^*$ is thereby obtained by first computing the centered heat flux tensor

\begin{equation}
    \bar{Q}_{\alpha\beta\gamma} = \sum_{i=1}^n f_i (\xi_{i\alpha}-u_\alpha) ( \xi_{i\beta}-u_\beta) (\xi_{i\gamma}-u_\gamma).
\end{equation}
Then, the nonequilibrium part $\bar{Q}_{\alpha\beta\gamma}^\mathrm{neq} = \bar{Q}_{\alpha\beta\gamma} - \bar{Q}_{\alpha\beta\gamma}^\mathrm{eq}$ is applied to the Hermite tensor $\mathcal{H}_{i\alpha \beta \gamma}^{(3)}$ 
\begin{equation}
    h^*_i = w_i \bar{Q}_{\alpha\beta\gamma} ^\mathrm{neq} : \mathcal{H}_{i\alpha \beta \gamma}^{(3)},
\end{equation}
via full contraction of indices.

\bibliography{PRE}

\end{document}